# Direct Validation of Superconductivity through Contact-Free Detection of Persistent Supercurrents Using Room-Temperature Quantum Magnetometry

Xinyi Zeng [a, b], Chengzhen Qin [b,c], Bowen Fan [d], Hammad Ahmed [a], Hui Fang [e], Stuart Long [a], Xiaonan Shan [a], Jiefu Chen [a], Shoujun Xu [f], Liangzi Deng [b,g] , Ching W. Chu [b,g] and Jiming Bao [a,b,f,g,*]

[a] Department of Electrical & Computer Engineering, University of Houston, Houston, Texas 77204, USA

[b] Texas Center for Superconductivity (TcSUH), University of Houston, Houston, Texas 77204, USA

[c] Materials Engineering Program, University of Houston, Houston, Texas 77204, USA

[d] Department of Mechanical and Aerospace Engineering, University of Houston, Houston, Texas 77204, USA

[e] Department of Physics, Sam Houston State University, Huntsville, Texas 77340, USA

[f] Department of Chemistry, University of Houston, Houston, Texas 77204, USA

[g] Department of Physics, University of Houston, Houston, Texas 77204, USA

*To whom correspondence should be addressed. Email: jbao@uh.edu.

# Abstract

The accelerated emergence of new materials is driving the search for high temperature superconductors, but rapid experimental validation remains a critical bottleneck, particularly for microscopic samples under high pressure. Here, we demonstrate one-step direct superconductivity validation through room-temperature, contact-free detection of remnant supercurrents. The technique utilizes a cryogen-free optically pumped atomic magnetometer to detect the temperature-dependent magnetic field produced by supercurrents of a superconductor. The superconducting transition is directly identified by the abrupt disappearance of magnetic field from the remnant supercurrent above the transition temperature and the reversal of the supercurrent direction upon reversing the applied magnetic field. Validated on YBCO microcrystals and REBCO tape, this technique detects pico-Tesla magnetic fields from supercurrents induced by the ambient Earth's magnetic field in a millimeter-sized REBCO square disk, as well as from sub-100 µm YBCO microcrystals compatible with high-pressure diamond anvil cells. The use of a ferrite flux guide enables sensitive detection from centimeter-scale distances. Requiring no electrical contacts, magnetic coils, or integrated magnetic sensors, this non-invasive, room-temperature platform offers a scalable approach for high-throughput screening and validation of superconductivity.

## Introduction

Recent years have witnessed a rapid increase in the number of candidate high-temperature and near-room-temperature superconductors, including nickelates, hydrides, LK-99, C–S–H and Lu–N–H [1-5], a trend expected to accelerate with the emergence of autonomous materials discovery enabled by artificial intelligence (AI), machine learning, high-throughput computation, and robotic synthesis [6-11]. As computational platforms predict an ever-growing number of promising superconductors, rapid and reliable experimental validation has become a critical bottleneck. The conventional hallmarks of superconductivity, i.e., zero electrical resistance and the Meissner effect, are often difficult to establish in newly synthesized materials because of impurity phases, microscopic sample dimensions, and extreme synthesis conditions. This challenge is particularly severe for hydride superconductors synthesized in diamond anvil cells (DACs), where samples are typically only tens of micrometers in size and the available experimental space is extremely limited [12-14]. Consequently, conventional transport and magnetic susceptibility measurements are technically demanding and can yield inconclusive or controversial results [5, 15-17], slowing the verification of new superconductors. The growing gap between AI-enabled materials discovery and experimental validation calls for a simple, reliable, and scalable technique for rapidly screening and validating superconductivity, particularly in microscopic samples synthesized under extreme conditions [18].

The defining property of a superconductor is its ability to sustain persistent supercurrents without electrical dissipation, arising from macroscopic quantum coherence. Both zero electrical resistance and the Meissner effect are manifestations of this underlying quantum state, while remnant supercurrents form the physical basis of superconducting technologies ranging from MRI and NMR magnets to particle accelerators, fusion magnets, and magnetically levitated transportation.

Rather than separately measuring two consequences of superconductivity, direct detection of persistent supercurrents probes the superconducting state itself and therefore provides a more fundamental and compelling validation [19]. This concept has recently attracted renewed interest for high-pressure hydride superconductors through magnetic detection of trapped flux using superconducting quantum interference devices (SQUIDs) [20]. Although SQUIDs provide exceptional magnetic sensitivity [21, 22], they require cryogenic operation, expensive instrumentation, and customized sample holders or DAC configurations, limiting their practicality and widespread use. In addition, SQUIDs detect changes in magnetic flux rather than the absolute magnetic field generated by the remnant supercurrent, making quantitative measurements on microscopic samples challenging. Mechanical motion of the DAC during measurement can further introduce magnetic noise and compromise measurement reliability. These limitations motivate the development of a simple, room-temperature, contact-free method capable of directly detecting persistent supercurrents with high sensitivity.

Here we introduce Direct Superconductivity Validation (DSV), a contact-free quantum magnetometry platform that validates superconductivity through direct detection of persistent remnant supercurrents. DSV employs a room-temperature optically pumped atomic magnetometer to directly measure the absolute magnetic field generated by remnant supercurrents [23, 24], eliminating the need for electrical contacts, susceptibility coils, cryogenic sensors, or integrated magnetic probes. Compared with SQUID-based approaches [21, 22], DSV greatly simplifies the experimental configuration, reduces cost, and is particularly well suited for superconductors with transition temperatures near or above liquid-nitrogen temperature. We demonstrate DSV using YBCO microcrystals and a commercial REBCO superconducting tape [13, 25], and show that magnetic fields generated by superconducting samples smaller than 100 μm, comparable in size to hydride superconductors synthesized in diamond anvil cells, can be detected from distances of approximately 2 cm using a ferrite flux guide. Beyond providing a robust and unambiguous validation of superconductivity, DSV offers a practical platform for rapid screening of large numbers of candidate materials emerging from AI-driven materials discovery.

## Results

The principle and experimental procedure of DSV are illustrated in Fig. 1. When initial temperature, $T_{ini}$, is below its superconducting transition temperature, $Tc$, a superconductor can sustain persistent supercurrents that generate a stable magnetic field, allowing it to function as a permanent electromagnet (Fig. 1a). As the temperature increases from $T_{ini}$ and rises above $Tc$, the superconducting state collapses, causing the persistent supercurrent and its associated magnetic field to disappear abruptly (Fig. 1b). Monitoring the magnetic field generated by the persistent supercurrent therefore provides a direct validation of superconductivity while simultaneously identifying the superconducting transition temperature.

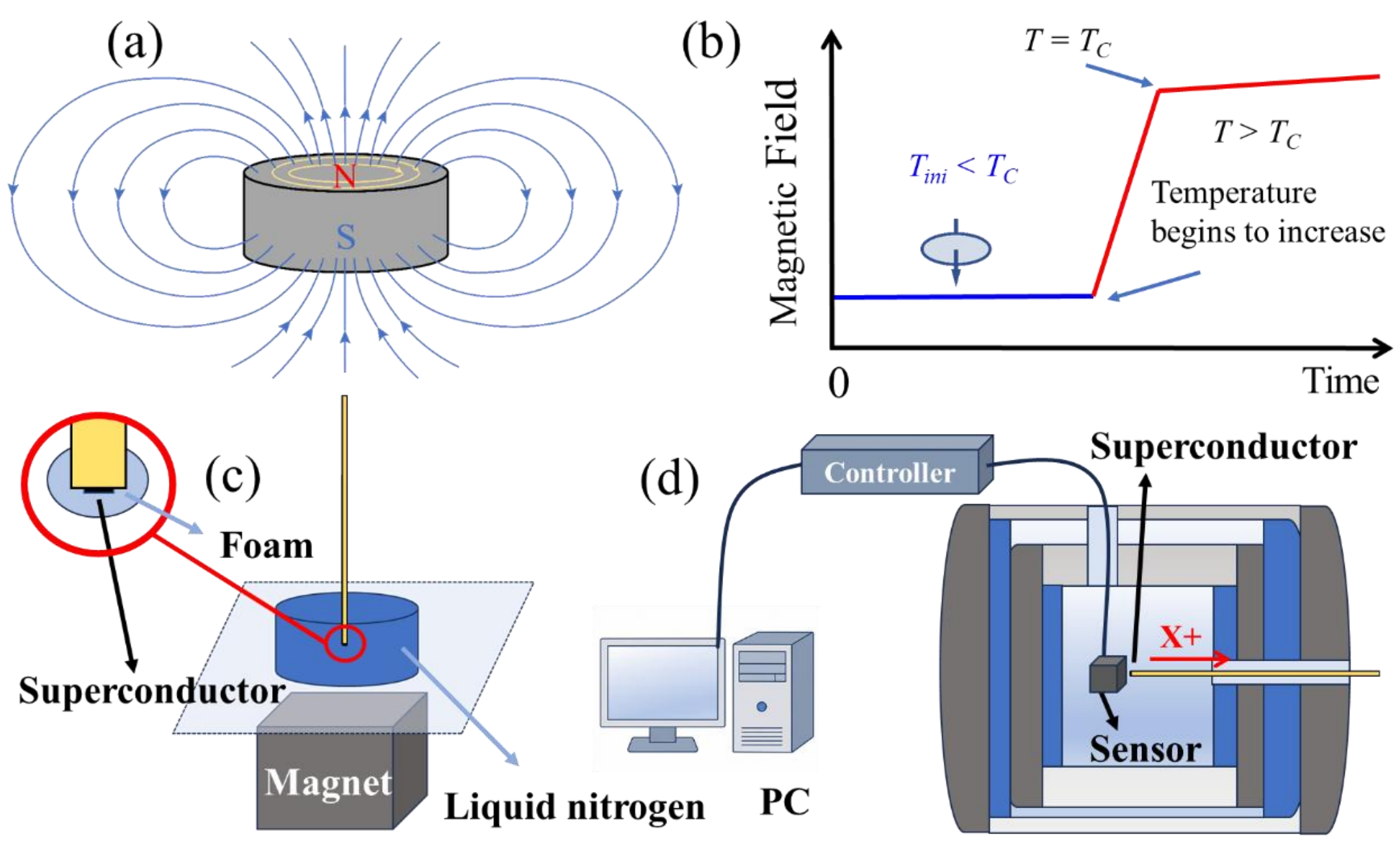

Figure 1. Principle of Direct Superconductivity Validation (DSV). (a) Below the superconducting transition temperature (*Tc*), a superconductor sustains persistent supercurrents. (b) As the temperature rises above *Tc* from initial $T_{ini}$ < *Tc*, the persistent supercurrent disappears abruptly, causing the magnetic field to collapse. (c) Field-cooling procedure used to induce persistent supercurrents in a superconductor with *Tc* < 77 K. (d) Contact-free detection of the magnetic field generated by the remnant supercurrent using an optically pumped atomic magnetometer inside a magnetically shielded chamber at room-temperature.

Figures 1c and 1d illustrate the experimental configuration and measurement procedure. To demonstrate DSV, we investigate two well-established high-temperature superconductors: YBCO microcrystals [13, 25] and a commercial REBCO superconducting tape (Shanghai Superconductor Technology Co., Ltd.), both with transition temperatures above 77 K, the boiling point of liquid nitrogen. A superconducting sample is attached to the tip of a nonmagnetic support (a wooden cotton swab) and field-cooled in liquid nitrogen above a NdFeB permanent magnet. The applied magnetic field is conveniently controlled by adjusting the distance between the magnet and the sample [26]. After 1–2 min of field cooling, the sample is removed from the liquid nitrogen and immediately transferred into a magnetically shielded chamber containing a highly sensitive optically pumped atomic magnetometer. After the residual liquid nitrogen is rapidly evaporated, the sample warms continuously under ambient conditions. When the temperature exceeds *Tc*, the persistent supercurrent vanishes, producing a corresponding abrupt collapse of the measured magnetic field, as illustrated in Fig. 1b. This abrupt transition constitutes the defining signature of DSV and provides a simple, contact-free, and direct validation of superconductivity.

We employ a commercial quantum magnetometer: a QZFM Gen-3 zero-field magnetometer from QuSpin, which is an optically pumped atomic magnetometer based on a rubidium vapor cell.

Figure 2a shows the background signal of the magnetometer inside the magnetic shield, exhibiting a noise level of approximately 2–3 pT, demonstrating its high sensitivity. Figures 2b and 2d show the magnetic signal measured when the wooden stick of a cotton swab was manually inserted into the magnetic shield through an access port and then withdrawn after approximately 6 s. Although the wooden stick is nonmagnetic, its presence produced a background signal of about 70 pT. To test the accuracy of the magnetometer, we fabricated a circular copper coil consisting of 10 turns with a radius of 1.25 mm. As shown in Fig. 2e, the coil was attached directly to the sensor housing using adhesive tape. A direct current of 10 μA generated a measured magnetic field of 122 pT in the X− direction (Fig. 2c), in reasonable agreement with the calculated value of 163 pT based on the coil geometry. The remaining discrepancy is likely attributable to imperfect alignment between the coil axis and the sensor's sensitive axis, together with uncertainties in the exact coil geometry and the influence of stray magnetic fields from the wire connecting the coil.

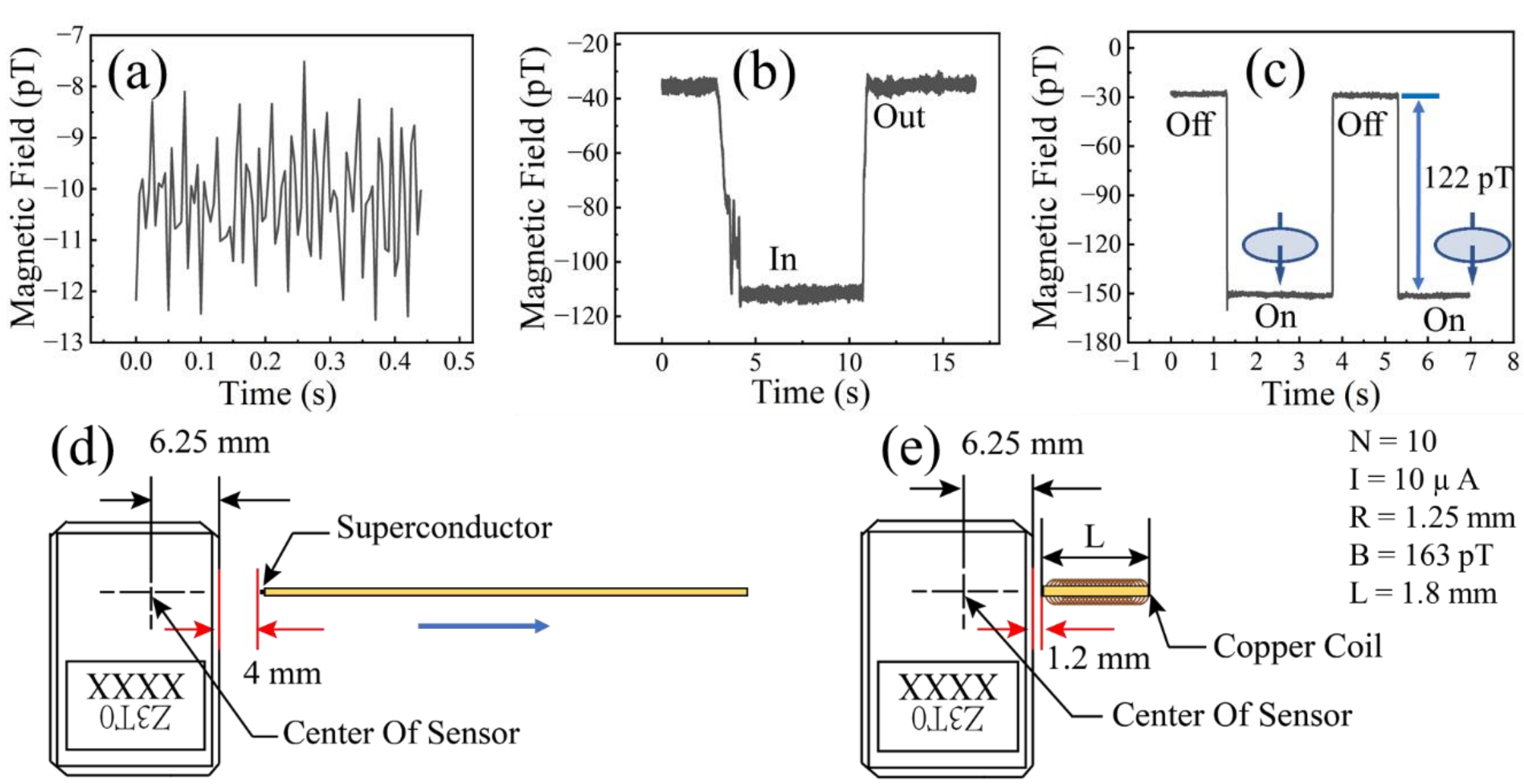

Figure 2. Sensitivity and test of the QuSpin atomic magnetometer. (a) Background magnetic field measured by the sensor inside the magnetic shield. (b) Magnetic

signal recorded when a wooden stick was inserted into and removed from the shield.
(c) Magnetic field generated by a 10-turn copper test coil carrying a 10 μA current.
(d, e) Experimental configurations showing the relative positions of the sensor and
(d) the wooden stick and (e) the test coil.

Having established the sensitivity and operation of the measurement system, we next measured the magnetic field generated by remnant supercurrents. Because superconducting samples synthesized in diamond anvil cells are typically only tens of micrometers in size, we first evaluated the technique using small YBCO chips obtained by mechanically breaking a bulk YBCO crystal. Figure 3a shows an optical image of a representative YBCO microcrystal. The sample was attached to the tip of the wooden stick using adhesive tape and covered with a small piece of foam to retain a small amount of liquid nitrogen. Following the procedure illustrated in Fig. 1, the sample was field-cooled in liquid nitrogen under an applied magnetic field for approximately one minute and then immediately transferred into the magnetic shield, where it remained stationary throughout the measurement. Figure 3b shows the evolution of the measured magnetic field. The magnetic field remained at ~50 pT before it began to increase in the X− direction at ~1 s, when the sample and stick were inserted into the magnetic shielding. The magnetic field reached ~−450 pT when the sample was fully positioned in the shielding chamber at ~2.5 s. The total field change was therefore approximately 500 pT, substantially larger than the ~70 pT background measured for the wooden stick alone (Fig. 2b), indicating a significant contribution from the remnant supercurrent in the YBCO sample.

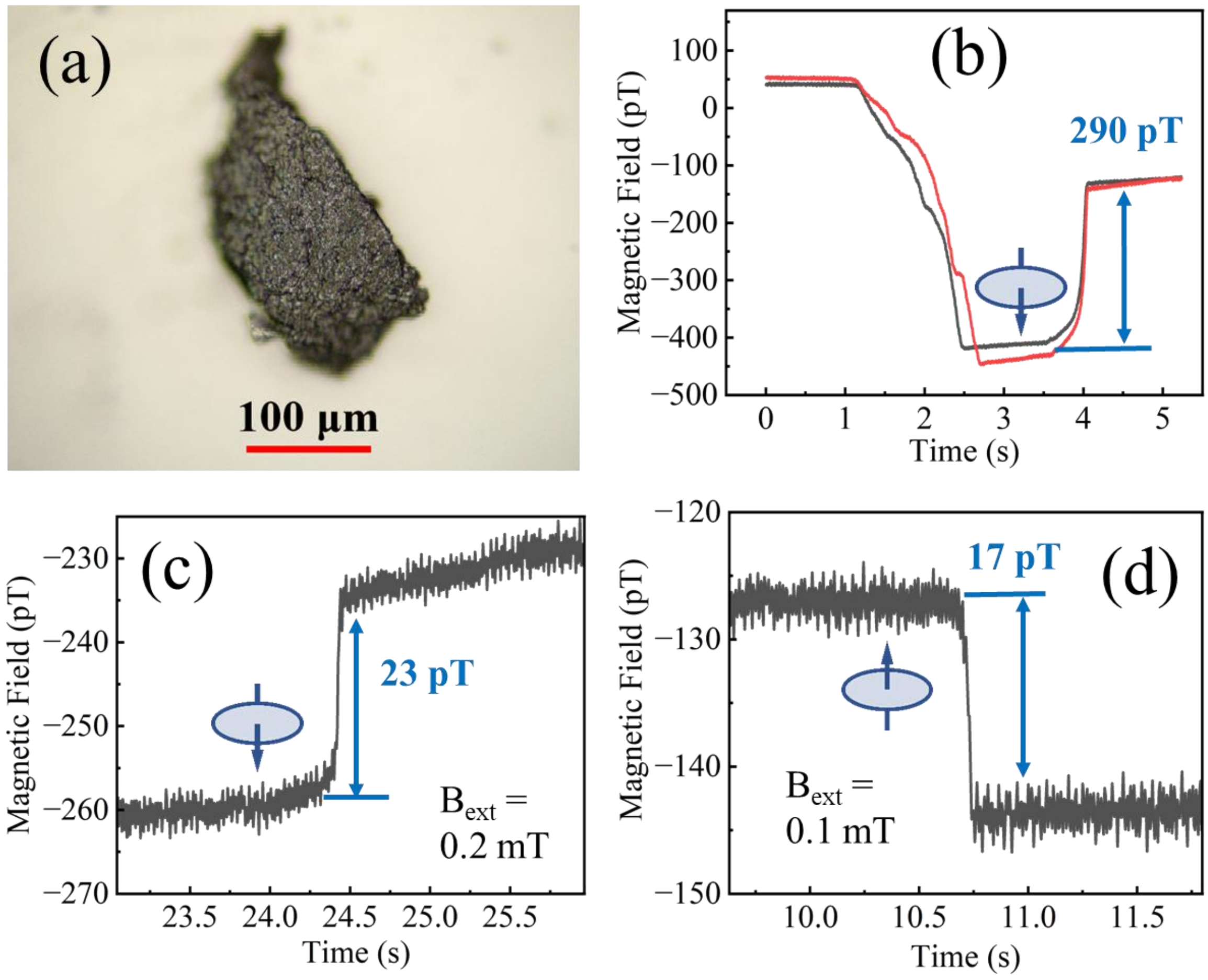

Figure 3. Detection of remnant supercurrent in a YBCO microcrystal. (a) Optical image of a YBCO chip broken from a bulk single crystal. (b) Time evolution of the magnetic field generated by the remnant supercurrent after field cooling under an applied magnetic field of 3.4 mT. (c, d) Magnetic fields after field cooling under weak external magnetic fields of opposite polarity, demonstrating the reversal of the remnant supercurrent direction.

The magnetic field remained nearly constant for approximately one second, after which it exhibited an abrupt decrease by 290 pT and returned to a weak background signal comparable to that of the bare wooden stick. This sudden collapse of the magnetic field is attributed to the disappearance of the remnant supercurrent as the sample temperature rapidly increased from 77 K to above *Tc* owing to the evaporation of the residual liquid nitrogen. Thus 290 pT was the magnetic field produced by the supercurrent in X− direction at 77 K. To further verify the superconducting origin of the measured magnetic field, we reversed the direction of the applied magnetic field by simply flipping the permanent magnet before field cooling. As shown in Figs. 3c and 3d, the measured magnetic field reversed its sign while exhibiting the same abrupt disappearance near the superconducting transition. This reversal confirms that the remnant supercurrent is determined by the polarity of the applied magnetic field and provides compelling evidence that the observed magnetic signal originates from persistent superconducting currents rather than magnetic contamination or other spurious effects. A ferromagnetic phase can be quickly excluded as it can be magnetized spontaneously below its Curie temperature without an external applied magnetic field.

To further evaluate the sensitivity of the sensor and establish the quantitative relationship between the applied magnetic field and the remnant supercurrent, we selected a smaller YBCO microcrystal and systematically varied the applied magnetic field during the field-cooling process. Figure 4a shows an optical image of the YBCO microcrystal used in this experiment. Figures 4b–e present the magnetic field generated by the remnant supercurrent after field cooling under four representative applied magnetic fields of 3.4, 7.4, 20.9 and 530 mT. Figure 4f summarizes the results obtained under all eight increasing magnetic fields. At low external magnetic fields, the average remnant magnetic field increases approximately linearly with the applied field [27]. As the applied field exceeds approximately 20 mT, the magnetic field, and consequently the remnant supercurrent, begins to saturate. This behavior is expected for type-II superconductors under field-cooling conditions [27]. Although the measured magnetic field exhibits considerable variation, primarily due to variations in the manual field-cooling and sample-transfer procedures, the overall trend is clear and reproducible.

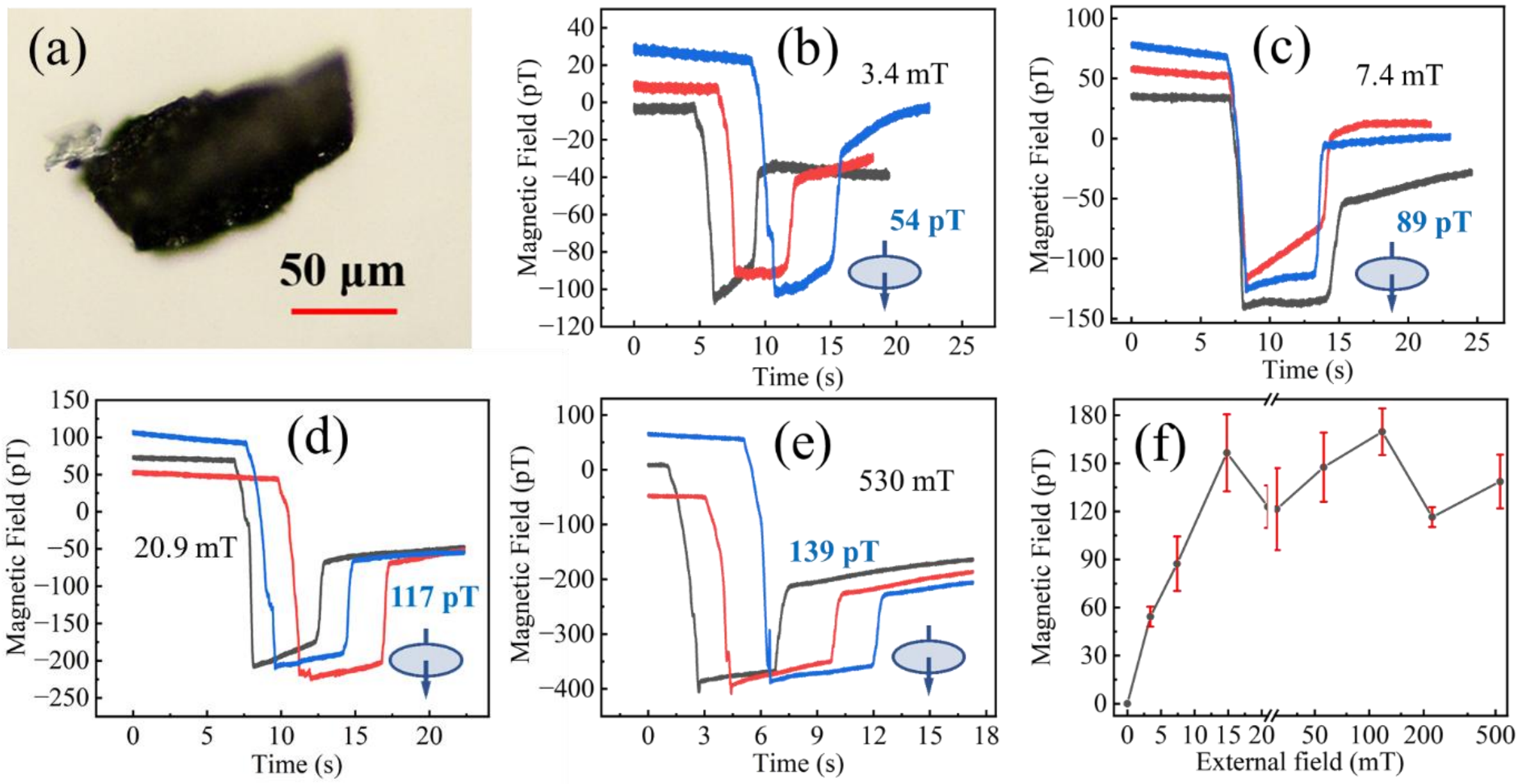

Figure 4. Dependence of the remnant supercurrent on the applied magnetic field for a smaller YBCO microcrystal. (a) Optical image. (b–e) Time evolution of the magnetic field generated by the remnant supercurrent after field cooling under four representative applied magnetic fields. (f) Average magnetic field generated by the remnant supercurrent as a function of the applied magnetic field.

The successful detection of remnant supercurrents in YBCO microcrystals motivated us to evaluate a commercially available REBCO superconducting tape. Because the superconducting layer is supported by multiple metallic substrate layers, the tape cannot be cleaved or broken as easily as a bulk YBCO crystal. Instead, we cut the tape into a square specimen with dimensions of approximately $1.3 \times 1.3$ mm$^2$. Figure 5a shows an optical image of the REBCO sample and Fig. 5b shows a schematic of its multi-layer structure. When the sample was field-cooled under a relatively large external magnetic field, the remnant supercurrent generated a magnetic field that exceeded the dynamic range of the magnetometer. Therefore, a smaller applied field of 4 mT was used. As shown in Fig. 5c, the remnant supercurrent produced a magnetic field of approximately 5.5 nT, more than an order of magnitude larger than that measured from the YBCO microcrystals. Remarkably, even when no external magnet was applied during field cooling, a measurable remnant magnetic field of approximately 20 pT was observed (Fig. 5d). We attribute this signal to supercurrents induced solely by the Earth's magnetic field during cooling, which marks the first observation of supercurrent induced by the ambient magnetic field.

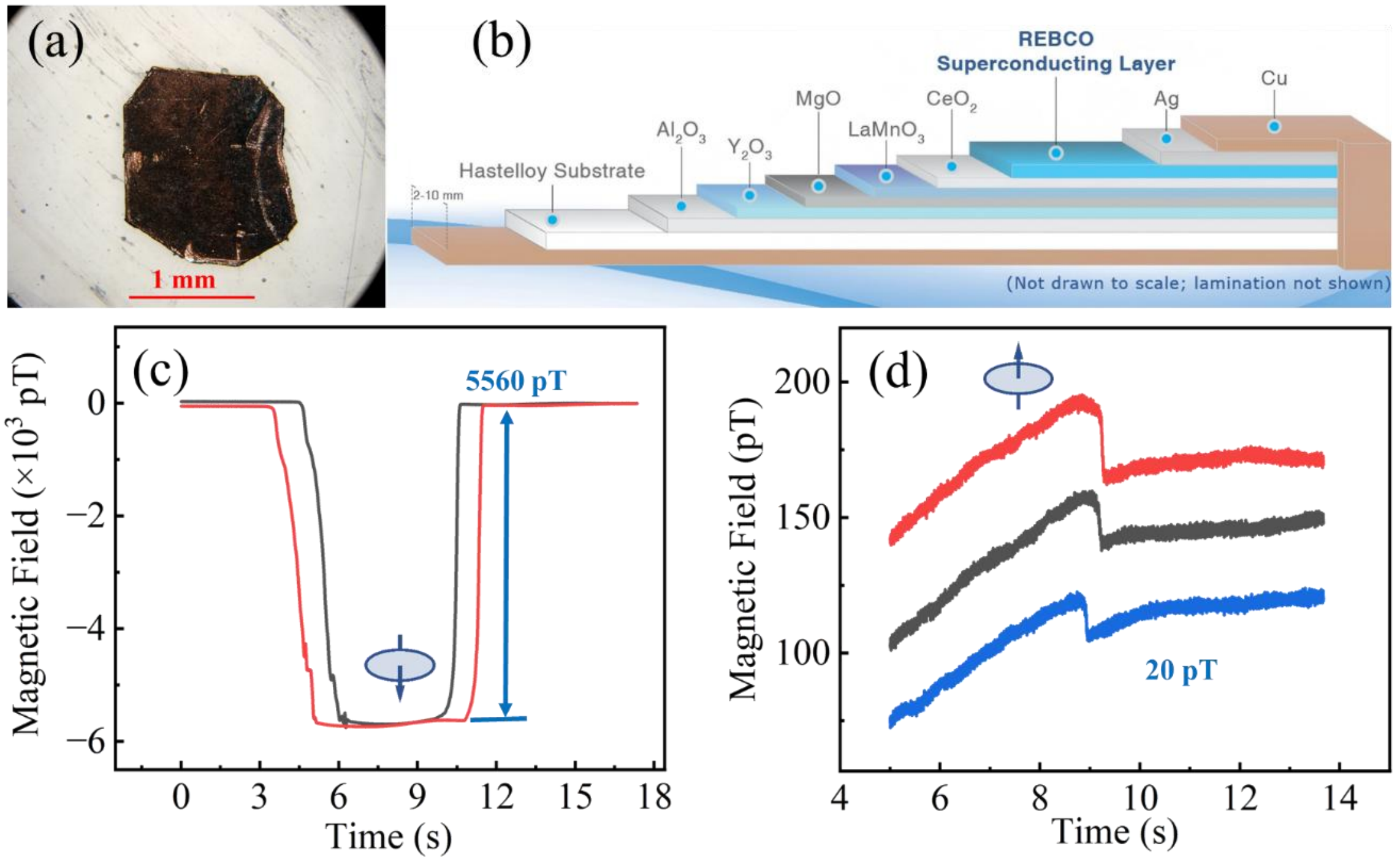

Figure 5. Detection of remnant supercurrents in a commercial REBCO superconducting tape. (a-b) Optical image and cross-section structure of a REBCO sample. (c) Magnetic field generated by the remnant supercurrent after field cooling under an applied magnetic field of 4 mT. (d) Remnant magnetic field generated after field cooling in the ambient Earth's magnetic field without an externally applied magnet.

In the experiments described above, the superconducting sample was positioned as close as possible to the quantum magnetometer to maximize the detected magnetic signal. In practical high-pressure experiments, however, the sample is enclosed within a DAC, which significantly increases the sensor-to-sample distance. A typical DAC has a length of approximately 3.6 cm; therefore, when the sample is located near the center of the cell, the minimum distance to an external magnetometer is approximately 1.8 cm. Because the dimensions of a superconducting sample are much smaller than this distance, the remnant supercurrent can be approximated as a magnetic dipole. Consequently, the magnetic field decreases approximately as $1/R^3$, where R is the distance between the sample and the sensor [28]. This rapid decay presents a major challenge for detecting the weak magnetic fields generated by microscopic superconductors inside a DAC and represents one of the principal obstacles to contact-free validation of high-pressure superconductivity.

To overcome this distance limitation, we introduce a magnetic flux concentrator that enables remote detection of remnant supercurrents [29]. The concentrator consists of a ferrite rod placed between the superconducting sample and the quantum magnetometer. Owing to its high magnetic permeability, the ferrite rod efficiently collects, concentrates, and guides the magnetic flux generated by the remnant supercurrent toward the sensor, thereby substantially enhancing the detectable magnetic signal without requiring electrical connections or active components. To demonstrate this concept, we compared three experimental configurations: a short sensor-to-sample distance, a large sensor-to-sample distance, and a large distance with a 15 mm ferrite rod inserted between the sample and the sensor (Figs. 6a–c). The corresponding magnetic field measurements are shown in Figs. 6d–f for the same YBCO microcrystal used in Fig. 3 after field cooling in a 5 mT magnetic field. Increasing the sensor-to-sample distance from approximately 6 mm to 21 mm reduced the measured magnetic field from about 630 pT to 90 pT, in good agreement with the expected $1/R^3$ dependence of a magnetic dipole. In contrast, inserting the ferrite flux guide restored the detected magnetic field to nearly its original magnitude, demonstrating its remarkable ability to guide magnetic flux over centimeter-scale distances with minimal signal loss. Because the magnetic flux concentrator is entirely passive and readily scalable, remote detection can be extended to even greater distances by employing longer ferrite rods or optimized flux-guiding geometries.

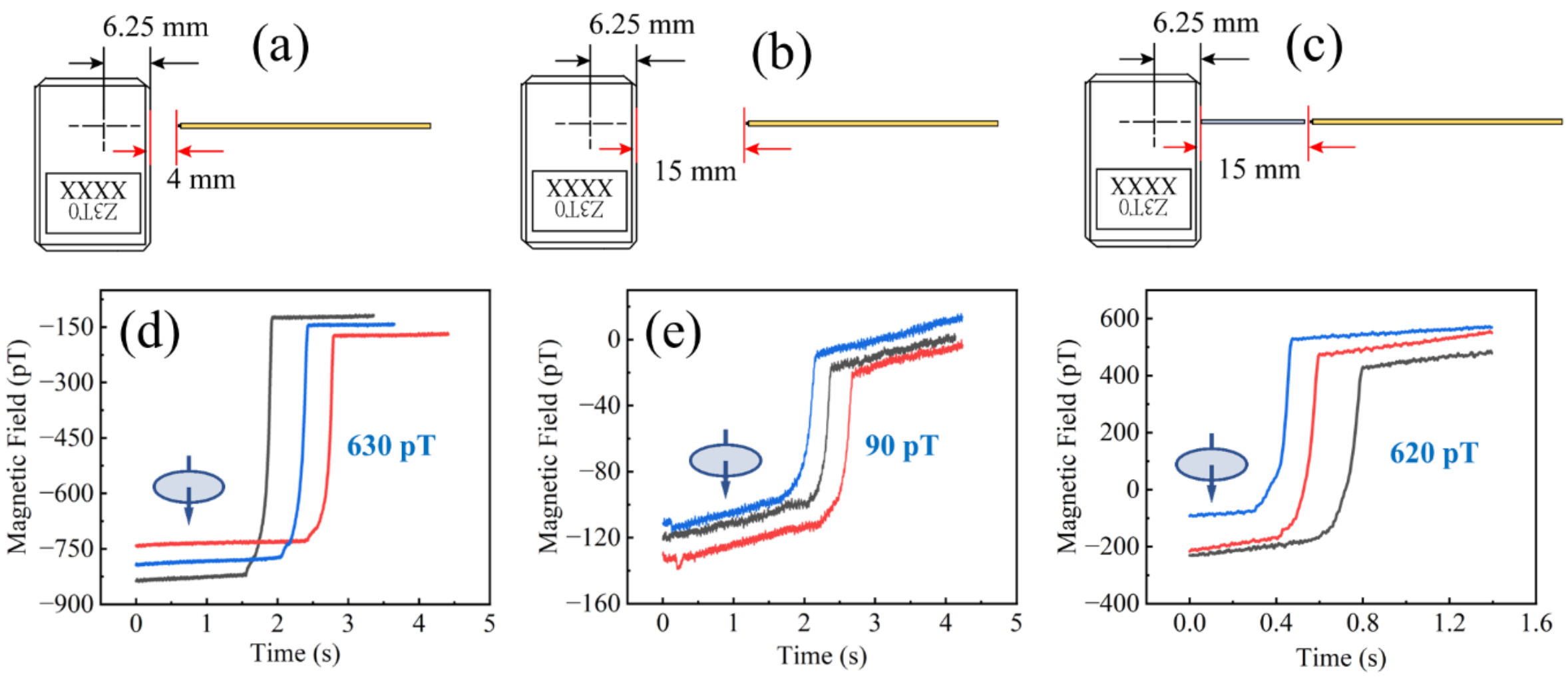

Figure 6. Distant detection of remnant magnetic fields using a ferrite flux guide. (a-c) Three experimental configurations. (a) At a short distance of ~10 mm between the sample and the sensor. (b) At a large distance of ~21 mm. (c) At a large distance of ~21 mm with a 15-mm long ferrite rod between the sample and the sensor. (d-f) Corresponding magnetic fields by the same YBCO sample as in Fig. 3 under 5 mT magnetic field cooling.

## Discussion

Although DSV is demonstrated here using well-established type-II superconductors at ambient pressure, the technique is broadly applicable to superconductors of different types and under a wide range of experimental conditions. For type-I superconductors, a persistent current loop can be readily created by introducing a microscale hole or non-superconducting region at the center of the sample using laser micromachining or mechanical drilling. For high-pressure superconductors, including metal hydrides synthesized in DACs, the compact room-temperature quantum magnetometer can readily accommodate a standard DAC within the magnetic shielding. When additional space is required, the shielding can be readily enlarged or custom fabricated using high-permeability μ-metal without affecting the operating principle of the technique.

DSV is equally applicable to superconductors with lower transition temperatures. By mounting the sample or DAC on a temperature-controlled cold finger or cryogenic stage, the superconducting transition, temperature dependence of the remnant supercurrent, and supercurrent relaxation dynamics can be investigated with high precision. Compared with conventional transport or SQUID magnetic measurements, DSV eliminates the need for microfabricated electrical contacts, miniature DACs, susceptibility coils, or integrated quantum sensors, greatly simplifying experiments on microscopic samples. This contact-free configuration is particularly advantageous for DAC experiments because it preserves valuable space around the sample for simultaneous laser synthesis, Raman spectroscopy, X-ray diffraction, optical spectroscopy, or other in situ characterization techniques without concern for laser damage to electrodes or quantum sensors [14]. Beyond superconductivity, the room-temperature quantum magnetometry platform can be readily extended to investigate the magnetic properties of other quantum materials, including two-dimensional magnets, antiferromagnets, ferrimagnets, and topological magnetic materials, by directly measuring their stray magnetic fields.

## Conclusion

In conclusion, we have introduced DSV, a room-temperature contact-free quantum magnetometry platform that directly validates superconductivity through detection of persistent remnant supercurrents. Unlike conventional approaches based on separate measurements of zero electrical resistance and the Meissner effect, DSV directly probes the underlying quantum phenomenon responsible for both signatures using an optically pumped atomic magnetometer, eliminating the need for electrical contacts, susceptibility coils, cryogenic magnetic sensors, or implanted quantum defects. We demonstrate pT-sensitive detection of remnant supercurrents from superconducting samples smaller than 100 μm and show that a passive magnetic flux concentrator enables remote measurements over a large distance. The simplicity, scalability, and compatibility of DSV with high-pressure synthesis and in situ materials characterization make it a practical platform for rapid superconductivity screening and validation. By transforming superconductivity validation from a technically demanding experiment into a simple, room-temperature, contact-free measurement, DSV establishes a new experimental paradigm for the AI era of superconducting materials discovery and paves the way toward the realization of ambient-pressure, room-temperature superconductors [18].

## Declaration of competing interest

SX holds equity in UForce Biotechnology, LLC. This financial interest has been disclosed to the University of Houston, and appropriate steps have been taken to address any potential conflicts of interest.

## Data availability

Data will be made available on request.

**Acknowledgements**

J.M.B. acknowledge support from the U.S. National Science Foundation under the award number DMR-2529884.